\documentclass[12pt]{iopart}
\usepackage{times}
\usepackage{mathptmx}
\usepackage{graphicx}

\begin{document}

\title[Defect correlations, MIT, and magnetic order
in ferromagnetic semiconductors]{Defect correlations,
metal-insulator transition, and magnetic order
in ferromagnetic semiconductors}
\author{Carsten Timm and Felix von Oppen}
\address{Freie Universit\"at Berlin, Arnimallee 14, D-14195 Berlin,
Germany}
\begin{abstract}
Diluted ferromagnetic III-V semiconductors typically show a high degree of
compensation. Compensation is connected to the presence of comparable
densities of charged defects of either sign. This naturally leads to the
development of strong correlations between defect positions during growth
and annealing. We show that these correlations are required to understand
the experimentally observed transport and magnetic properties as well as
the persistence of the energy gap upon doping with magnetic ions.
\end{abstract}

\section{Introduction}

Diluted ferromagnetic III-V semiconductors are promising materials for
applications as well as interesting from the physics point of view. They
might allow the incorporation of ferromagnetic elements into semiconductor
devices, and thus the integration of data processing and magnetic storage
on a single chip. It is important to understand the interplay between
transport properties, magnetic ordering, and the defect configuration,
which is determined by the growth conditions. We show that correlated
positions of defects are required for a description consistent with
experiments \cite{TSO}.

To be specific, we consider GaAs doped with manganese, but the results
should apply for most III-V compounds. The physics relies on the dual role
played by the Mn impurities: They carry a local spin due to the half-filled
\emph{d}-shell and dope the system with holes, which mediate a
ferromagnetic indirect exchange interaction between the spins. A crucial
feature of these materials is their high degree of compensation, which is
presumably due to antisite defects (As substituted for Ga). The accordingly
small density of holes leads to weak electronic screening of the Coulomb
interaction between charged defects and between defects and holes. This
role of compensation has not been taken into account previously. Due to the
strong Coulomb interactions a random distribution of defects on the cation
sublattice is very costly in energy. Therefore, defect diffusion
\cite{Potashnik} leads to a rearrangement of defects in such a way that the
Coulomb energy is reduced. We study this rearrangement with the help of
Monte Carlo (MC) simulations and find that strong correlations of defect
positions develop. The resulting disorder potential acting on the
valence-band holes is strongly reduced and most of its correlations fall
off on the scale of the minimum defect separation. We show that this effect
has tremendous consequences for transport properties, the integrity of the
energy gap, and the temperature-dependent magnetisation. In fact, a
description in agreement with experiments \emph{requires} the defects to be
strongly correlated.

\section{Correlated defects}

In this section, we study how diffusion of defects, \emph{i.e.}, Mn
impurities and antisites, changes the defect configuration and the disorder
potential. This is motivated by experiments \cite{Potashnik} that show that
defect diffusion is rather pronounced at typical growth and annealing
temperatures of the order of $250^\circ$C. We start from a random
distribution of Mn impuritites and antisites (As) on the cation sublattice,
where the density of antisites is determined by charge neutrality from the
observed \cite{Ohno} hole concentrations at given Mn doping levels. The
Hamiltonian reads
\begin{equation}
H = \frac{1}{2}\, \sum_{i,j} \frac{q_iq_j}{\epsilon r_{ij}}\,
  e^{-r_{ij}/r_{\mathrm{scr}}} ,
\end{equation}
where $q_i$ are the defect charges and $r_{ij}$ is their separation.
Relative to the cation sublattice, Mn impurities carry charge $q=-1$
whereas antisites carry $q=+2$.
The screening length $r_{\mathrm{scr}}$ is obtained from nonlinear
screening theory \cite{SE}. It is much larger than the
nearest-neighbor separation on the cation sublattice so that it
hardly affects the small-scale defect correlations relevant here. We perform
MC simulations for the Hamiltonian $H$ \cite{TSO} employing
the Metropolis algorithm at the temperature $250^\circ$C for systems of
$20\times20\times20$ conventional cubic unit cells with periodic boundary
conditions, unless stated otherwise. To roughly model the dynamics,
we only use local MC moves, \emph{i.e.}, exchanges of nearest-neighbor
defect and/or host atoms. A more realistic modelling of the dynamics
involving interstitials and vacancies lies beyond the scope of this work.
The resulting configurations are quenched at low
temperatures \cite{Keld}.

\begin{figure}[h]
\hspace*{2mm}
\includegraphics[width=8.7cm]{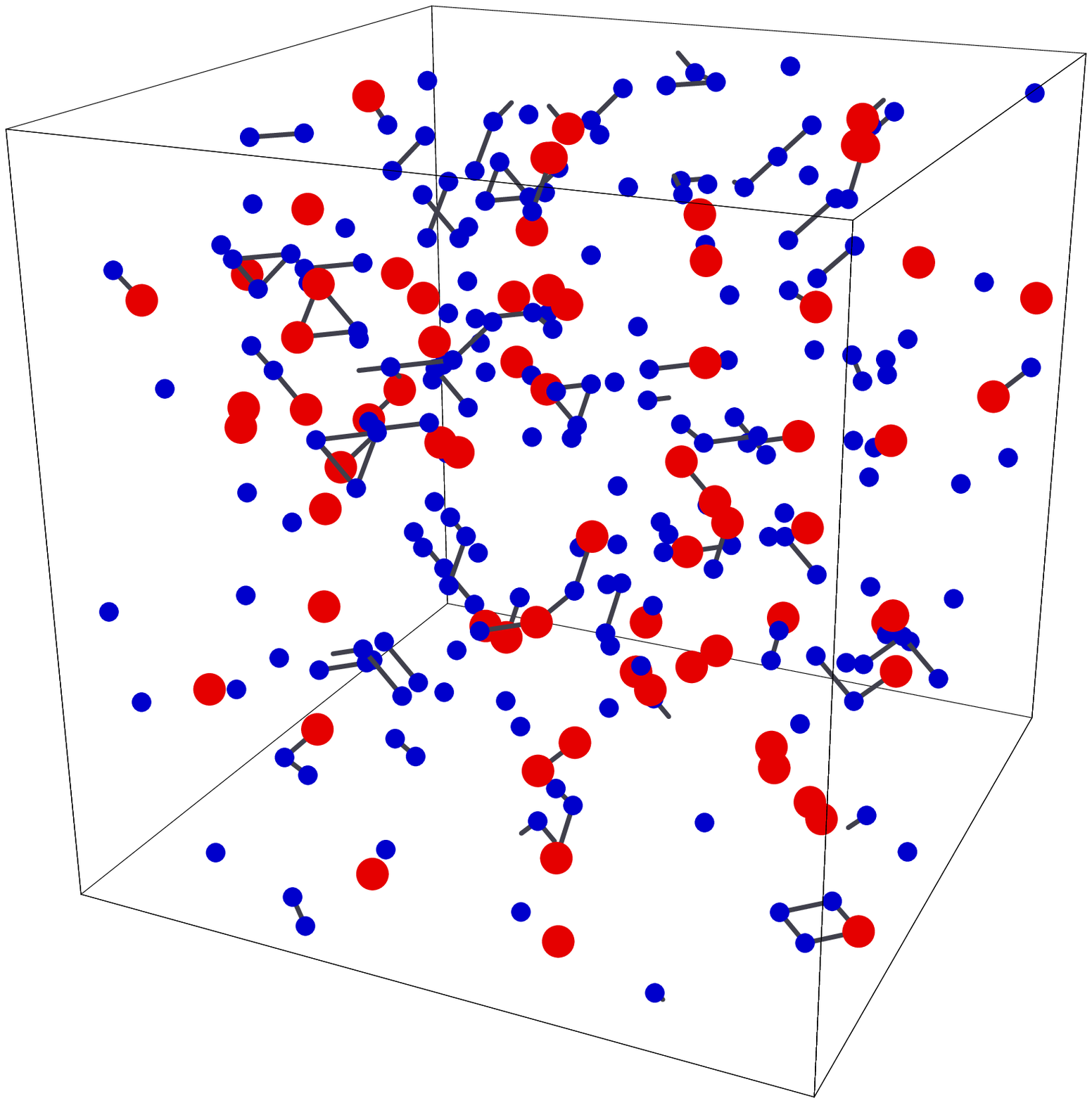}
\hspace*{-8mm}
\includegraphics[width=8.7cm]{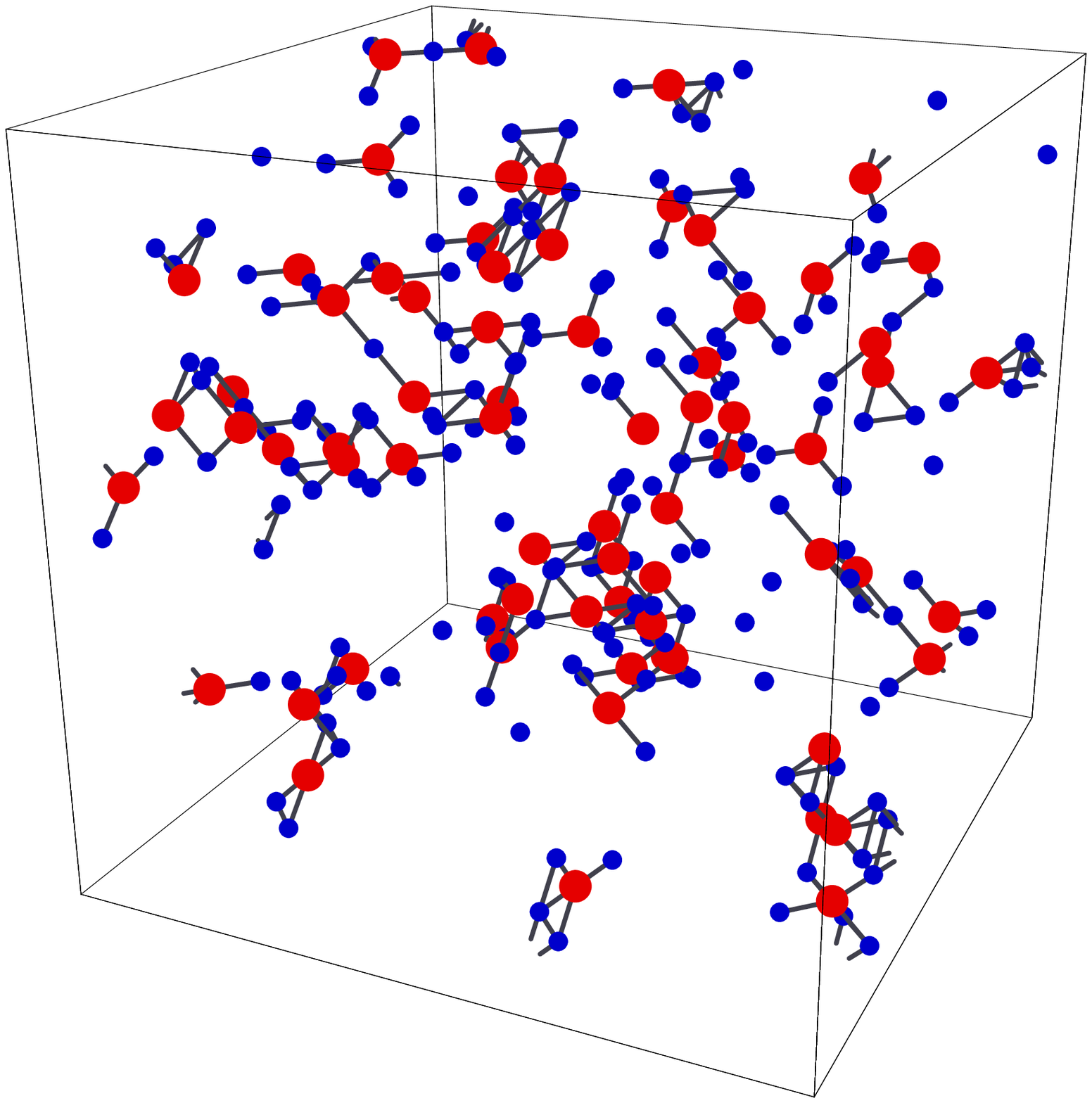}
\caption{\label{fig.conf0}Left: Random configuration of defects on a
$10\times10\times10$ lattice for Mn concentration $x=0.05$ and $p=0.3$ holes
per Mn. The host lattice is not shown.
Small blue circles denote Mn impurities, whereas large red ones are As
antisites. When two defects sit on nearest-neighbor sites, they are
connected by a solid line. Right: Equilibrated configuration of defects.
It is obvious that the defect positions are
now strongly correlated.}
\end{figure}

Figure \ref{fig.conf0} (left) shows a random configuration of defects for a
Mn concentration of $x=0.05$ and $p=0.3$ holes per Mn, which determines the
concentration of antisites ($0.0175$). This should be compared to the right
part of Fig.~\ref{fig.conf0}, which shows an equilibrated configuration of
the same number of defects. It is obvious that the defects have formed
small clusters, where typically an antisites (charge $+2$) is surrounded by
several Mn impurities (charge $-1$). It is clear that this clustering leads
to a strongly reduced impurity potential, since the antisite charges are
very effectively screened.

\begin{figure}[h]
\centerline{\includegraphics[width=11cm]{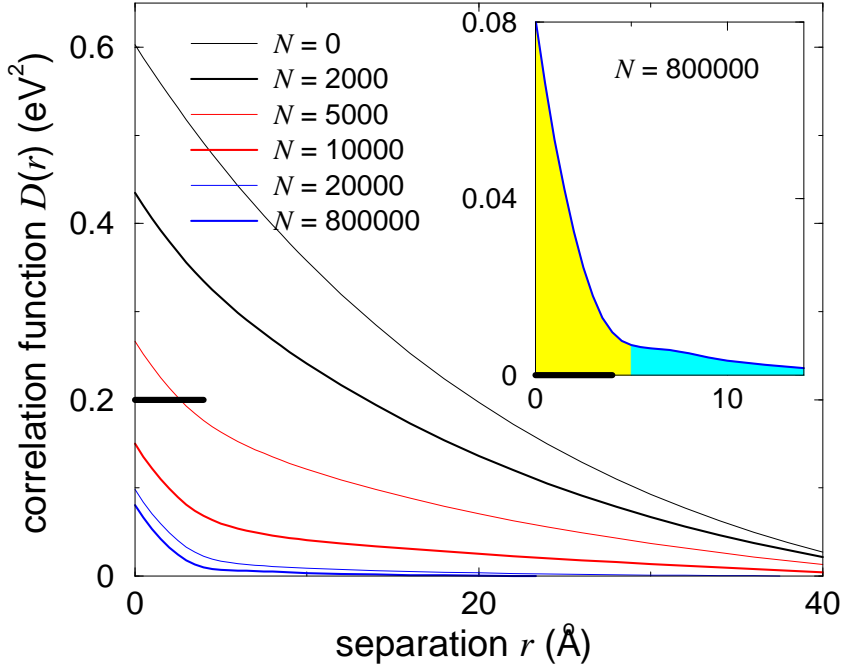}}
\caption{\label{fig.correl}Potential correlation function $D(r)$ for Mn
concentration $x=0.05$ and $p=0.3$ holes per Mn, plotted for various
numbers of MC steps. The heavy solid bar denotes the nearest-neighbor
separation on the cation sublattice. The inset shows the initial
rapid decrease of $D(r)$ for the equilibrated case, see text.}
\end{figure}

More quantitative information can be obtained from the
potential correlation function
\begin{equation}
D(r) \equiv \langle
   V({\bf r})\,V({\bf r}') \rangle_{\scriptscriptstyle |{\bf r}-{\bf r}'|=r} -
   \langle V \rangle^2 .
\end{equation}
Obviously, $\Delta V\equiv\sqrt{D(0)}$ is the width of the distribution of
$V(\mathbf{r})$. $D(r)$ is plotted in Fig.~\ref{fig.correl} for $x=0.05$
and $p=0.3$ for various numbers of MC steps, which serve as a rough
meassure of annealing time. Defect diffusion strongly reduces $D(r)$ and
thus $\Delta V$. The inset in Fig.~\ref{fig.correl} shows a blowup of
$D(r)$ for the approximately equilibrated state. The initial reduction of
$D(r)$ to about 10\% of $D(0)$ on the scale of the nearest-neighbor
separation of about $4.00$\AA (heavy bar) is due to the screening of the
compensated 70\% of Mn impurities by antisites, leaving only 30\% active.
Since $D(r)$ contains the potential squared, this leads to a reduction to
about 9\% (yellow region in the inset). The remaining contribution from
uncompensated Mn (light blue region) decays on the typical length scale of
$14.4$\AA associated with their density, which is $0.3\times 0.05$ per
cation site. Clearly, ionic screening, \emph{i.e.}, screening by the
defects, is nearly perfect at $250^\circ$C.
While the reduction of $\Delta V$ is thus substantial, we find that
$\Delta V$ is still \emph{not} small compared to the Fermi energy \cite{TSO}.

\section{Valence-band holes}

In this section we consider the properties of valence-band holes
\cite{comment} in the impurity potential $V(\mathbf{r})$ due to the charged
defects. In particular, we are interested in their spectrum and
localisation properties. We employ the envelope function and parabolic-band
approximations for the holes and start from the Hamiltonian
$H=-\sum_i(\hbar^2/2m^\ast)\,\nabla_i^2+V(\mathbf{r}_i)$. For
material-specific calculations the detailed band structure should be taken
into account, \emph{e.g.}, using the 6-band Kohn-Luttinger Hamiltonian
\cite{KL}. This should not change the qualitative results but is known to
increase the mean-field $T_c$ \cite{KL}. The hole Hamiltonian is written in
a plane-wave basis and diagonalized numerically \cite{TSO}. The
calculations are done for spin-less holes, since the additional disorder
introduced by the exchange interaction is found to be much smaller than
$\Delta V$. We obtain the energy spectrum and normalized eigenfunctions
$\psi_n({\bf r})$. From the latter we calculate the inverse participation
ratios
\begin{equation}
\mathrm{IPR}(n)=1/\sum_{\bf r} |\psi_n({\bf r})|^4
\end{equation}
of the states. The main physical content of the IPR is that it scales with
system size for extended states but essentially remains constant for
localized states. The IPR thus allows to estimate the position of the
mobility edge in the valence band.

\begin{figure}[h]
\centerline{\includegraphics[width=11cm]{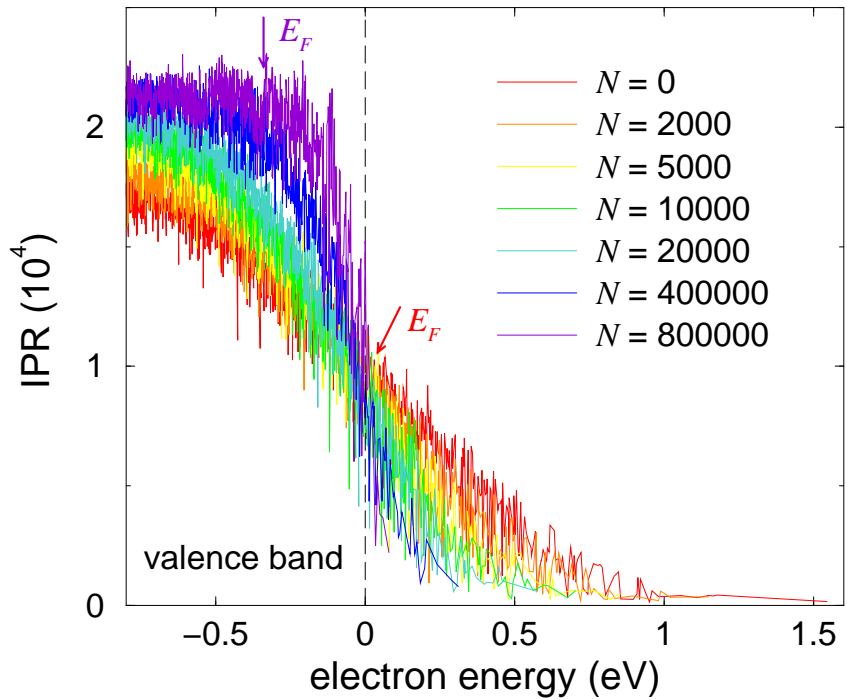}}
\caption{\label{fig.ipr1}IPR as a function of electron energy for
$x=0.05$ and $p=0.3$ after various numbers of MC steps increasing from the
red to the violet curve. The Fermi energies for the random and the
equilibrated case are also shown.}
\end{figure}

Figure \ref{fig.ipr1} shows the IPR as a function of electron energy for
the same parameters as above, $x=0.05$ and $p=0.3$, after various numbers
of MC steps. The plot shows that uncorrelated defects would lead to the
filling-in of the band gap by disorder, which is in contradiction to
experiments. Thus correlated defects are required to explain the
persistence of the energy gap.
We have also studied the IPR for various system sizes
and find that states in the flat region and the upper part of the slope are
extended, whereas states in the band tail are localized, as expected.
Figure \ref{fig.ipr1} shows that the
states at the Fermi energy become more and more extended with annealing,
since the disorder potential decreases.

\begin{figure}
\centerline{\includegraphics[width=11cm]{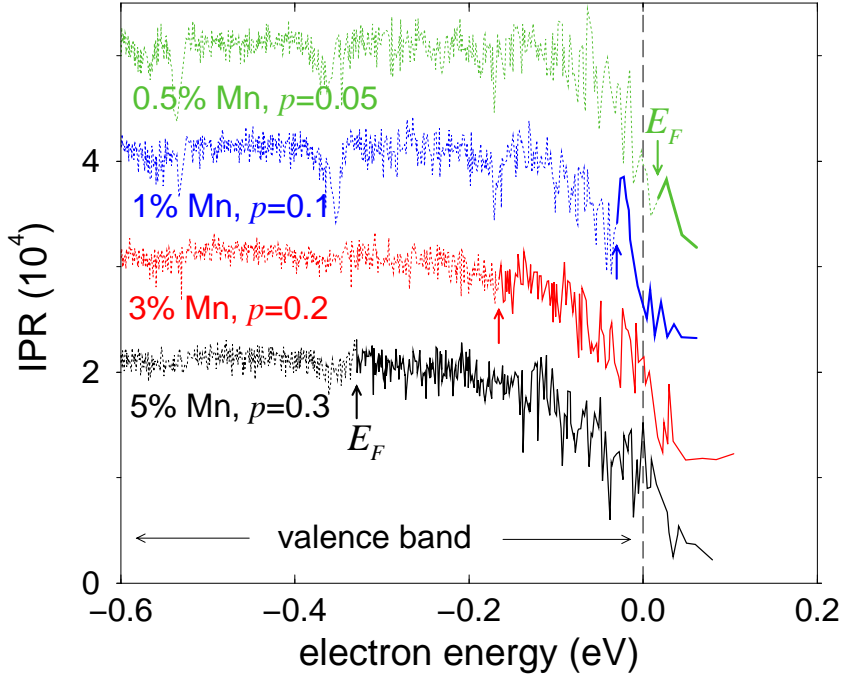}}
\caption{\label{fig.ipr2}IPR as a function of energy for equilibrated
configurations as a
function of Mn concentration $x$ with the number of holes per Mn, $p$,
chosen in accordance with experiments \protect\cite{Ohno}. The Fermi
energies are also indicated.}
\end{figure}

In Fig.~\ref{fig.ipr2} the IPR for fully annealed configurations is plotted
as a function of Mn concentration $x$. The number of holes per Mn, $p$, has
been chosen in accordance with experiments \cite{Ohno}. The change in the
IPR-curves with $x$ is rather weak due to the strong ionic screening. The
main effect is the shift of the Fermi energy due to the varying hole
density, which drives the Fermi edge into localized states below about 1\%
Mn. This qualitatively explains the metal-insulator transition observed in
(Ga,Mn)As \cite{Ohno}.

\section{Ferromagnetism}

Finally, we turn to the magnetic properties. The bulk magnetisation is
composed of contributions of the local Mn spins and of the holes. Since the
exchange interaction between valence-band holes and local Mn spins is
antiferromagnetic, these contributions are oriented oppositely. However,
the hole contribution turns out to be very small. We study the
magnetisation within a selfconsitent mean-field approximation, which we
briefly outline in the following.

The starting point is the Hamiltonian
of the coupled valence-band holes and Mn spins,
\begin{equation}
H = \sum_{n\sigma} \xi_n\, c_{n\sigma}^\dagger c_{n\sigma}
  - J_{\mathrm{pd}} \sum_i \mathbf{s}_i \cdot \mathbf{S}_i ,
\label{H10}
\end{equation}
where $\mathbf{S}_i$ are Mn spins ($S=5/2$) and
\begin{equation}
\mathbf{s}_i \equiv \sum_{n\sigma n'\sigma'} c_{n\sigma}^\dagger
  \psi_n^\ast(\mathbf{R}_i)\,
  \frac{\mbox{\boldmath$\sigma$}_{\sigma\sigma'}}{2}\,
  \psi_{n'}(\mathbf{R}_i) c_{n'\sigma'}
\end{equation}
are hole spin polarisations at the Mn sites. $\xi_n$ and $\psi_n$ are
the hole eigenenergies and eigenfunctions, respectively, for vanishing
magnetic interaction, $J_{\mathrm{pd}}=0$. These have been
obtained in the previous section---note that $H$ contains the full Coulomb
disorder potential. $\xi_n$ includes the hole chemical potential.

We rewrite Eq.~(\ref{H10}) as
\begin{equation}
H = \sum_{n\sigma} \xi_n\, c_{n\sigma}^\dagger c_{n\sigma}
  - \frac{J_{\mathrm{pd}}}{4} \sum_i \left[(\mathbf{s}_i+\mathbf{S}_i)^2 -
  (\mathbf{s}_i-\mathbf{S}_i)^2\right]
\end{equation}
and employ the imaginary-time functional integral. The two
spin-squared terms are decoupled using Hubbard-Stratonovich
transformations with auxilliary fields $\mathbf{h}_i^\pm$ coupling to
$\mathbf{s}_i\pm\mathbf{S}_i$. The auxilliary fields are then transformed
to new fields
$\mathbf{h}_i \equiv \mathbf{h}_i^+ + i \mathbf{h}_i^-$,
$\mathbf{h}'_i \equiv \mathbf{h}_i^+ - i \mathbf{h}_i^-$.
This leads to the Lagrangian
\begin{equation}
L = \sum_{n\sigma} c_{n\sigma}^\ast (\partial_\tau+\xi_n) c_{n\sigma}
  + i S\sum_i \dot\phi_i\,\cos\theta_i
  + \sum_i \frac{\mathbf{h}_i\cdot\mathbf{h}'_i}{J_{\mathrm{pd}}}
  - \sum_i (\mathbf{h}_i\cdot\mathbf{s}_i + \mathbf{h}'_i\cdot\mathbf{S}_i)
  .
\end{equation}
To obtain the mean-field theory we employ a saddle-point approximation. We
assume $\mathbf{h}_i$ and $\mathbf{h}'_i$ to be constant in time but
\emph{not} in space, since we wish to retain the disorder effects. The
holes and Mn spins are then easily integrated out. For the Mn spins this is
best done in the Hamiltonian formalism. This leads to the grand potential
\begin{equation}
\beta \Omega_0 = -\Tr \ln \beta G^{-1}(i\omega)
  - \sum_i \ln \frac{\sinh [\beta h'_i (S+1/2)]}{\sinh (\beta h'_i/2)}
  + \frac{\beta}{J_{\mathrm{pd}}} \sum_i \mathbf{h}_i \cdot \mathbf{h}'_i
\end{equation}
with the mean-field hole Green function
\begin{equation}
G^{-1}_{n\sigma n'\sigma'}(i\omega) =
  (-i\omega+\xi_n)\,\delta_{nn'}\delta_{\sigma\sigma'}
  - \sum_i \psi_n^\ast(\mathbf{R}_i)\,
  \frac{\mbox{\boldmath$\sigma$}_{\sigma\sigma'}}{2}\cdot
  \mathbf{h}_i\, \psi_{n'}(\mathbf{R}_i) .
\label{G10}
\end{equation}
The saddle-point equations are obtained by setting the derivatives of
$\beta\Omega_0$ with respect to $\mathbf{h}_i$ and $\mathbf{h}'_i$ to zero,
\begin{eqnarray}
\mathbf{h}_i & = & J_{\mathrm{pd}} \mathbf{M}_i = J_{\mathrm{pd}}
  \frac{\mathbf{h}'_i}{h'_i}\, S\, B_S(\beta S h'_i) ,
\label{sp10a} \\
\mathbf{h}'_i & = & J_{\mathrm{pd}} \mathbf{m}_i = - \frac{J_{\mathrm{pd}}}{\beta}\,
  \sum_{i\omega} \sum_{n\sigma n'\sigma'}
  G_{n\sigma n'\sigma'}(i\omega)\, \psi_{n'}^\ast(\mathbf{R}_i)\,
  \frac{\mbox{\boldmath$\sigma$}_{\sigma'\sigma}}{2} \psi_n(\mathbf{R}_i)
\label{sp10b}
\end{eqnarray}
with the usual Brillouin function $B_S(x)$. To actually evaluate
Eq.~(\ref{sp10b}) we diagonalize the hole sector of the Hamiltonian or,
equivalently, the inverse Green function $G^{-1}$, writing $G^{-1} = {\cal
O} g^{-1} {\cal O}^\dagger$, where $g^{-1}_{n\sigma} = -i\omega +
\tilde\xi_{n\sigma}$ is diagonal. We assume the the mean-field Mn spin
polarisation $\mathbf{M}_i$ to be collinear and choose the $z$ axis along
$\mathbf{M}_i$. Then $G^{-1}$ is symmetric and ${\cal O}$ is orthogonal.
Also, $G^{-1}$ is then already diagonal in spin space so that we can
diagonalize separately in each spin sector. We denote the components of
${\cal O}$ for spin sector $\sigma$ by ${\cal O}_{nn';\sigma}$. The
inversion of $G$ is now trivial, $G = {\cal O} g {\cal O}^\dagger$. The
frequency sum can finally be performed, since ${\cal O}$ does not depend on
frequency. Introducing the new eigenfunctions
\begin{equation}
\tilde\psi_{n\sigma}(\mathbf{R}_i) = \sum_{n'} \psi_{n'}(\mathbf{R}_i)\,
  {\cal O}_{n'n;\sigma} ,
\label{tlpsi5}
\end{equation}
Equation (\ref{sp10b}) then obtains the form
\begin{equation}
\mathbf{h}'_i = J_{\mathrm{pd}} \sum_{n\sigma} n_F(\tilde\xi_{n\sigma})\,
  \tilde\psi_{n\sigma}^\ast(\mathbf{R}_i)\,
  \frac{\mbox{\boldmath$\sigma$}_{\sigma\sigma}}{2}\,
  \tilde\psi_{n\sigma}(\mathbf{R}_i) .
\label{sp10bb}
\end{equation}
The mean-field solution is obtained by iterating Eqs.~(\ref{sp10a}) and
(\ref{sp10bb}).

\begin{figure}
\centerline{\includegraphics[width=11cm]{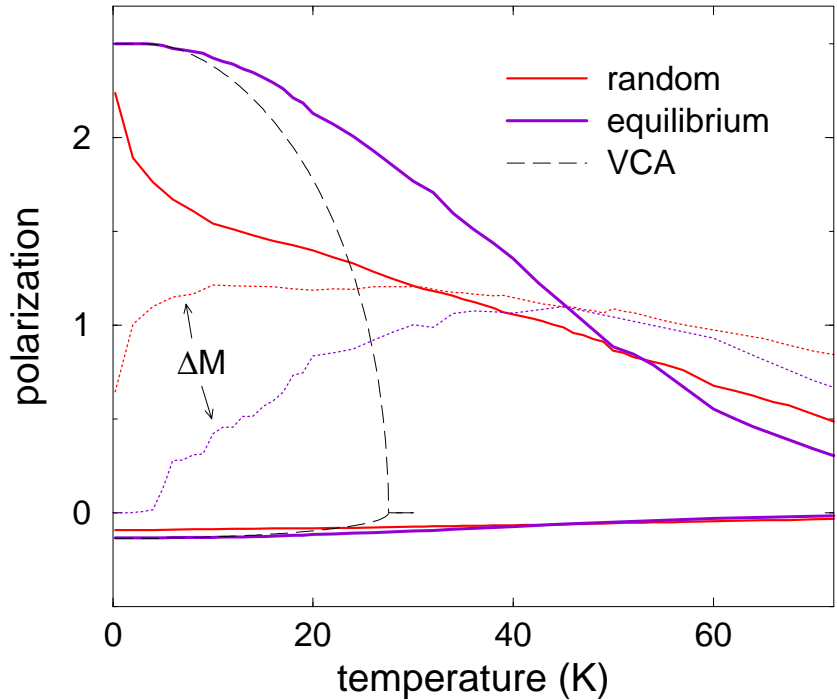}}
\caption{\label{fig.magn}Magnetisation of Mn spins
(up) and hole spins (down) as functions of temperature for
$x=0.05$ and $p=0.3$ for random, partially
annealed, and equilibrated configurations. The dotted curves give the
standard deviations $\Delta M$ of Mn spins.
For comparison, the long-dashed curves show the
magnetisations obtained from a theory that totally neglects disorder.}
\end{figure}

Figure \ref{fig.magn} shows the magnetisation curves for Mn and hole spins,
again for $x=0.05$ and $p=0.3$ for random and
equilibrium configurations. The hole-spin contribution to the total
magnetisation is found to be very small. The shape of the Mn magnetisation
curve becomes more ``normal'' and concave from below with annealing
\cite{TSO}, except for a tail at higher temperatures, in
good agreement with experimental annealing studies \cite{Potashnik}. The
anomalous shape for random defects is due to the localisation tendency of
the holes, which leads to a shorter-range effective Mn spin interaction.
This, together with the random positions of spins, leads to a broad
distribution of effective fields acting on these spins. For higher
temperatures, only a few spins with strong interactions carry most of the
magnetisation, leading to the drawn out tail for random defects. The
noise apparent on the curves is due to there being several saddle-points with
similar free energy and total magnetisation. We also plot the standard
deviations $\Delta M$ of the Mn spin polarisations. The
standard deviation becomes comparable with the average Mn spin polarisation
at intermediate temperatures below the Curie temperature $T_c$. This shows
that the reduction of the total magnetisation with increasing temperature
is initially dominated by disordering of large local moments. Only at
higher temperatures the local moments also decrease.

The Curie temperatures are significantly larger than what one obtains from
a virtual crystal approximation (VCA), neglecting disorder completely, see
Fig.~\ref{fig.magn}. It is an important open question whether this is an
artifact of mean-field theory or indeed a physical result. However, the
tails of the magnetisation curves with small average Mn spin polarisation
but large standard deviation are dominated by a few anomalously strongly
coupled spins. These are overemphasized by mean-field theory (remember that
mean-field theory gives an incorrect finite polarisation even for two
coupled spins). Thermal fluctuations should be effective in destroying the
long-range order in this case, thus reducing $T_c$.

\begin{figure}
\centerline{\includegraphics[width=11cm]{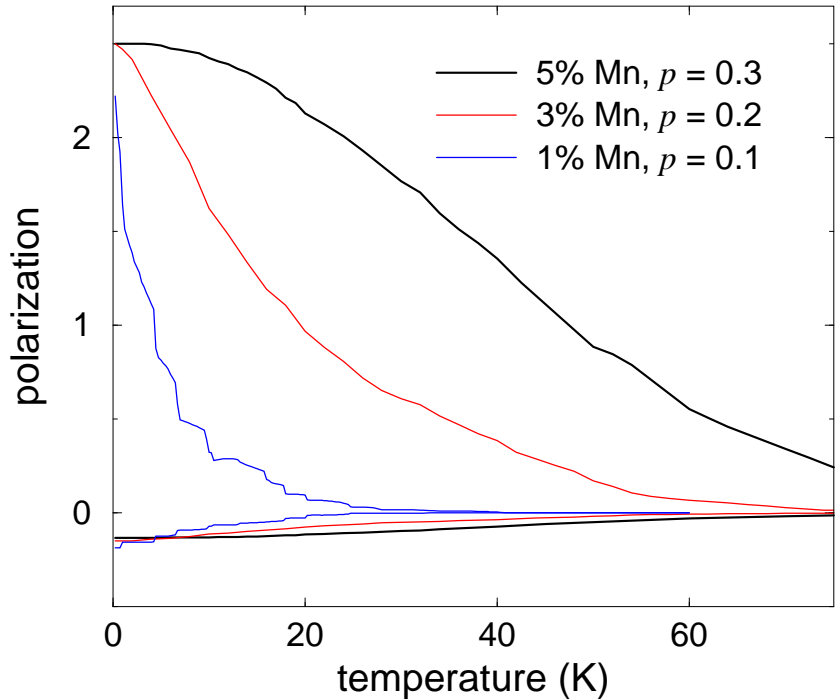}}
\caption{\label{fig.magn2}Magnetisation of Mn spins
(up) and hole spins (down) as functions of temperature for
various Mn and hole concentrations.}
\end{figure}

Finally, Fig.~\ref{fig.magn2} shows magnetisation curves for equilibrated
configurations at various Mn concentrations with the number of holes chosen
as above. We see that the curves for smaller Mn and hole concentrations
become more similar to the random limit in Fig.~\ref{fig.magn}, which is
easily understood since the holes become more localized.

\section{Conclusions}

We have shown that strong Coulomb interactions in highly compensated
diluted ferromagnetic semiconductors naturally lead to strong correlations
in the positions of charged defects. We also found that these correlations
are essential for the description of transport properties, in particular
the metal-insulator transition, the persistence of the energy gap, and the
shape of the magnetisation curves. It is a pleasure to acknowledge
discussions with P. J. Jensen, J. K\"onig, U. Krey, W. Nolting, M. E.
Raikh, F. Sch\"afer, and J. Schliemann.

\end{document}